\documentclass[aps,twocolumn,showpacs,preprintnumbers,amsmath,amssymb]{revtex4}
\usepackage{graphicx}
\usepackage{dcolumn}
\usepackage{bm}
\usepackage{hyperref}

\newcommand{\eqnref}[1]{Eq.~(\ref{#1})}
\newcommand{\figref}[1]{Fig.~\ref{#1}}

\begin{document}
\title{Quantum anomalous Hall effect on star lattice\\ with spin-orbit coupling and exchange field}
\author{Mengsu Chen}
\author{Shaolong Wan}
\altaffiliation{Corresponding author}
\email{slwan@ustc.edu.cn}
\affiliation{Institute for Theoretical Physics and Department of
Modern Physics, University of Science and Technology of China,
Hefei, 230026, P.R.China}

\date{\today}

\begin{abstract}
We study the star lattice with Rashba spin-orbit coupling and
exchange field and find that there exists the quantum anomalous
Hall effect in this system and there are five energy gaps at Dirac
points and quadratic band crossing points. We calculate the Berry
curvature distribution and obtain the Hall conductivity (Chern
number $\nu$) quantized as integers, and find that
$\nu=-1,2,1,1,2$, respectively, when the Fermi level lies in these
five gaps. Our model can be view as a general quantum anomalous
Hall system and, in limit cases, can give what the honeycomb
lattice and kagome lattice gave. We also find there exists a
nearly flat band with $\nu=1$ which may provide an opportunity to
realize the fractional quantum anomalous Hall effect. Finally, the
chiral edge states on a zigzag star lattice are given numerically
to confirm the topological property of this system.
\end{abstract}


\maketitle


{\it Introduction}.--The quantum Hall effect was observed
\cite{Klitzing1980} in 1980 in two dimensional electron system, in
which Hall conductivity takes quantized value
$\sigma_{xy}=\nu\,e^2/h$. The integer value $\nu$ is called the
TKNN number \cite{Thouless1982} or Chern number. The essential
ingredient of Hall effect is to break the time-reversal symmetry
of the system. Thus, introducing an external magnetic field is not
the only way to produce this effect. In fact, anomalous Hall
conductivity had been observed in ferromagnetic iron and
ferromagnetic conductors since 1881 \cite{Nagaosa2010}. The
internal magnetization plays an essential role in this so called
the anomalous Hall effect. It's naturally to ask if it is possible
to produce the quantum anomalous Hall (QAH) effect without
external magnetic field.

A model constructed by Haldane demonstrated that integral quantum
Hall (IQH) effect can be realized without the Landau level induced
by the external magnetic field \cite{Haldane1988}. In this
spinless model on the honeycomb lattice, staggered magnetic field
was introduced to break the time-reversal symmetry, while the
magnetic flux per unit cell is zero so that no Landau levels
present. This staggered magnetic field, in other words, the
complex next nearest hopping amplitude opens a gap at the Dirac
point. At half filling, when the Fermi level lies in this gap, the
Chern number of this system takes the value $\nu=\pm 1$. Besides
this toy model, several more realistic models which are based on
various systems, have been proposed. For example, in Anderson
insulators \cite{Nagaosa2003}, HgMnTe quantum well \cite{Liu2008},
optical lattices \cite{CjWu2008,YpZhang2011}, magnetic topological
insulators \cite{Yu2010TI}, graphene \cite{ZhQiao2010,WkTse2011},
kagome lattices \cite{ZyZhang2011}.

\begin{figure}
\begin{center}
\includegraphics[width=8.5cm]{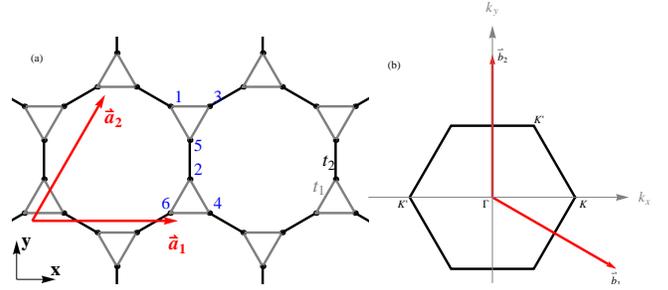}
\end{center}
\caption{(a) The lattice structure of the star lattice, where the
bond length is $a$ and $\vec{a_1}$, $\vec{a_2}$ are the primitive
vectors of its Bravais lattice. The sites in the unit cell are
labelled as $1$ to $6$. The nearest hopping amplitude within
triangles (all gray bonds) is $t_1$, the nearest hopping amplitude
between triangles (all black bonds) is $t_2$. (b) The reciprocal
vectors ($b_1$ and $b_2$) and the first Brillouin zone.}
\label{fig1}
\end{figure}

In this letter, we consider a star lattice model [see
\figref{fig1}(a)] which has close geometry connection to the
honeycomb lattice and kagome lattice \cite{Fiete2010} and study
its QAH effect. We calculate the Berry curvature in momentum space
and find the Hall conductivity quantized as an integer value when
the Fermi level lies in the topological nontrivial gap opened by
Rashba spin-orbital coupling and exchange field. Unlike the
proposals on honeycomb lattice and kagome lattice
\cite{ZhQiao2010,WkTse2011,ZyZhang2011}, five gaps are opened due
to the lattice structure complexity of the star lattice. Thus, in
our model, there exist five Hall plateaus when the Fermi level
lies in these gaps and there is a nearly flat band, with nonzero
Chern number ($\nu=1$), which may provide an opportunity to
realize the fractional quantum anomalous Hall (FQAH) effect when
electrons interaction is introduced. We also calculate on the
zigzag ribbon and demonstrate that the excitation states are
exactly local on the edge when the Fermi level lies in the bulk
gap.

{\it Hamiltonian}.--The star lattice can be seen as replacing the
site of honeycomb lattice by triangle, or replacing the site of
kagome lattice by segment. The number of sites per unit cell in
star lattice is six, three times as in honeycomb and twice as in
kagome. In this letter, we focus on the nearest hopping term and
assume that the hopping amplitude takes the same value $t_1$ in
triangle and $t_2$ between triangles (\figref{fig1}). The nearest
hopping tight-binding Hamiltonian is
\begin{equation}
H_0=-t_1\sum_{<i,j>,\alpha,\triangle}c_{i\alpha}^{\dag}c_{j\alpha}
-t_2\sum_{<i,j>,\alpha,\triangle\leftrightarrow\triangle}c_{i\alpha}^{\dag}c_{j\alpha}
+h.c., \label{1 tight binding Hamiltonian}
\end{equation}
where $\alpha$ is the spin index, $\triangle$ means "within
triangle" (all the gray bonds in \figref{fig1}a),
$\triangle\leftrightarrow\triangle$ means "between triangles" (all
the black bonds).

\begin{figure}
\begin{center}
\includegraphics[width=8.5cm]{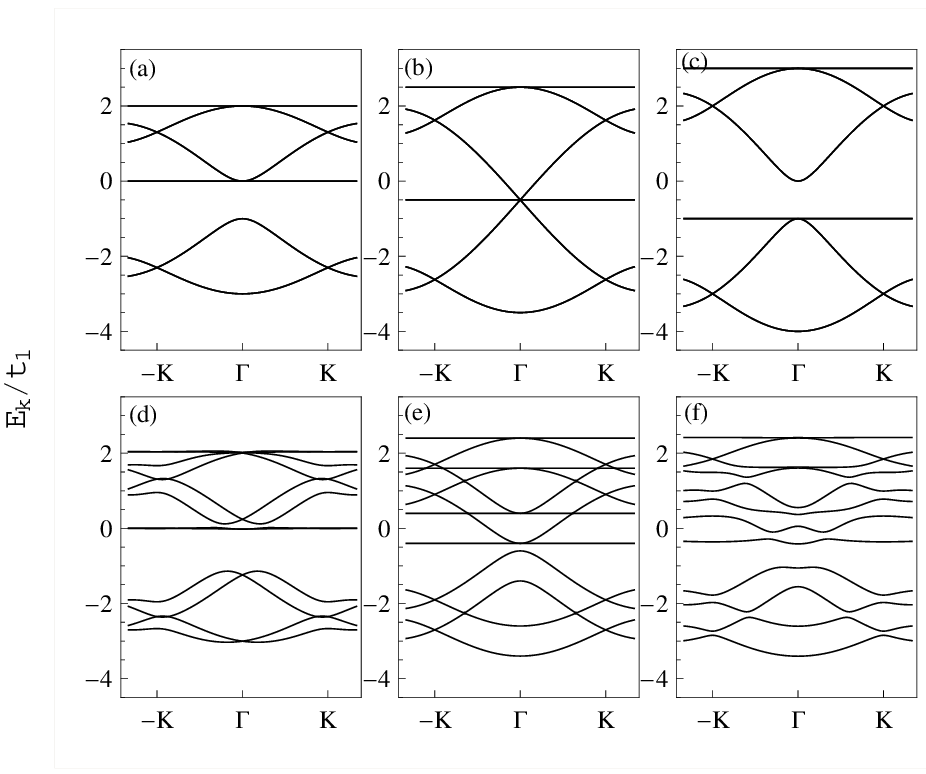}
\end{center}
\caption{(a) The band structures for the nearest hopping
tight-binding Hamiltonian $H_0$ with $t_2=t_1$. (b)
$t_2=\frac{3}{2}t_1$. (c) $t_2=2 t_1$. (d-f) The band evolution
for introducing Rashba spin-orbit coupling and exchange field when
$t_2=t_1$. (d) Only $H_{RSO}$ is introduced, the strength of
spin-orbit coupling is taken to be $t_{RSO}=0.2 t_1$. (e) Only
$H_{\lambda}$ is introduced, the strength of the exchange field is
taken to be $\lambda=0.2 t_1$. (f) Both $H_{RSO}$ and
$H_{\lambda}$ are introduced. The parameters are taken to be
$t_2=t_1$, $t_{RSO}=0.2t_1$, $\lambda=0.2 t_1$.} \label{fig2}
\end{figure}

By writing this Hamiltonian in momentum space and diagonalizing
the Hamiltonian matrix $H\left( k \right)$, the band structures
are obtained, as shown in \figref{fig2}(a)-(c). These band
structures demonstrate the close connections among the star
lattice, honeycomb lattice and kagome lattice. When $ t_2 <
\frac{3}{2} t_1$, the probability for electrons to hop out of the
triangle is smaller than that between triangles. This means that
the three points are bound together strongly and can be seen as
shrinking into one point. As a result, the low energy bands at
this case look like graphene [see \figref{fig2}(a)]. In contrast,
when $t_2 > \frac{3}{2} t_1$, the low bands look like kagome
lattice [see \figref{fig2}(c)]. We also notice that at
$t_2=\frac{3}{2}t_1$, the gap closes.

As shown in \figref{fig2}(a), a band gap already exists when only
considering the nearest hopping. However, this is not a
topological gap. In order to open topological gaps, we introduce
Rashba spin-orbit coupling and exchange field as follow
\begin{subequations}
\begin{align}
&H_{RSO}=it_{RSO}\sum_{<i,j>,\alpha,\beta}(\vec{\sigma}_{\alpha\beta}
\times \hat{d}_{ij})c_{i\alpha}^{\dag}c_{j\beta}+h.c., \\
&H_{\lambda}=\lambda\sum_{i,\alpha}c_{i\alpha}^{\dag}c_{i\alpha}\sigma^z_{\alpha
\alpha}, \label{2 KMGList1}
\end{align}
\end{subequations}
where $t_{RSO}$ is the strength of the spin-orbit coupling,
$\vec{\sigma}$ is the vector Pauli matrices in spin space,
$\hat{d}_{ij}$ is the unit vector point from site $j$ to $i$, and
$\lambda$ is the strength of the exchange field.

So, the total Hamiltonian is
\begin{equation}
H=H_0+H_{RSO}+H_{\lambda}. \label{3 TotalHamiltonian}
\end{equation}

The band evolution of this Hamiltonian is shown in \figref{fig2}.
Without Rashba spin-orbit coupling and exchange field, the
six-sites unit cell forms six bands which are doubly degenerated.
With only Rashba spin-orbit coupling $H_{RSO}$ is added, the spin
degeneracy is lifted, except at several $k$ points. The exchange
field alone takes the similar effect, but leaves degeneracy at
different $k$ points. And band gaps can be opened when both
interactions are added into the hopping Hamiltonian. In
following, we will confirm that these band gaps are topologically
nontrivial and that they can realize the quantum anomalous Hall
effect.

{\it Berry Curvature and Chern number}.--The Berry curvature in
crystal momentum space is an essential concept and should be
included when considering solid state system \cite{Xiao2010}.
Here, the intrinsic Hall conductivity can be written as the
summation of Berry curvature of all bands under Fermi level
\cite{Haldane2004} as follow
\begin{equation}
\sigma_{xy}=\frac{e^2}{h}\frac{2\pi}{N \mathcal{V}}\sum_{k,E_n\le
E_F}\Omega_z(E_n,k), \label{4 HallConductivity}
\end{equation}
where $N$ is the number of primitive unit cells, $\mathcal{V}$ is
the volume of the unit cell. The $n$th band's Berry curvature
$\Omega_z(E_n,k)$ can be obtained by
\begin{align}
&\Omega_z(E_n,k) \nonumber \\
&=\sum_{E_m(\neq E_n)}\frac{-2\text{Im}\langle\psi_{nk}|\partial
H\left(k\right)/\partial k_x|\psi_{mk}\rangle
\langle\psi_{mk}|\partial H\left(k\right)/\partial
k_y|\psi_{nk}\rangle}{(E_n-E_m)^2}. \label{5 band's Berry
curvature}
\end{align}
The Chern number is given by
\begin{equation}
\nu=\frac{2\pi}{N \mathcal{V}}\sum_{k,E_n(\leq
E_F)}\Omega_z\left(E_n,k\right), \label{6 Chern number}
\end{equation}
or
\begin{equation}
\nu=\frac{1}{2\pi}\sum_{E_n(\leq E_F)}\iint_{BZ} d^2
k\Omega_z\left(E_n,k\right), \label{7 Chern number}
\end{equation}
when $N\to\infty$.

\begin{figure}
\begin{center}
\includegraphics[width=8.5cm]{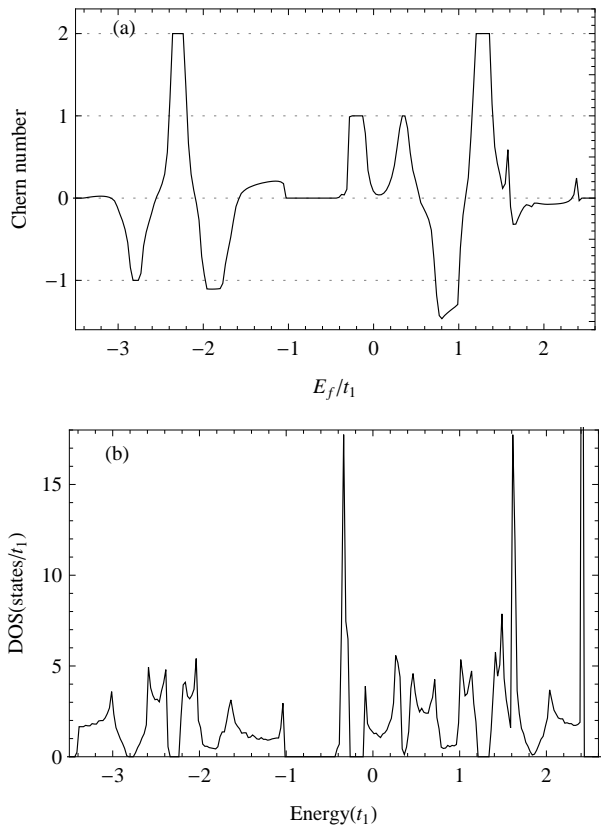}
\end{center}
\caption{(a) The Hall conductivity (or Chern number) as a function
of Fermi level $E_f$ with $t_2=t_1$, $t_{RSO}=0.2t_1$,
$\lambda=0.2 t_1$. (b) The corresponding density of states
(DOS).}\label{fig3}
\end{figure}

Writing the Hamiltonian (\ref{3 TotalHamiltonian}) in momentum
space and calculating the Hall conductivity by \eqnref{4
HallConductivity}, we get the Hall conductivity $\sigma_{xy}$ as a
function of Fermi level $E_F$, as shown in \figref{fig3}(a). Five
Hall plateaus are obtained when Fermi level lies in the gaps (as
suggested by the density of states shown in \figref{fig3}(b)). The
Chern number take quantized value $\nu=-1,2,1,1,2$, respectively.

The two $\nu=2$ plateaus come from the gaps opened at Dirac
points, which are similar to that in honeycomb lattice
\cite{ZhQiao2010} and kagome lattice \cite{ZyZhang2011}. The two
$\nu=1$ plateaus come from the gaps opened at quadratic band
crossing points (QBCP), which are similar to the kagome lattice
\cite{ZyZhang2011}. However, there is only one $\nu=1$ plateau in
kagome lattice. What's more, one of these nonzero Chern number
belongs to the nearly flat band. This novel property provides the
great potential to realize the fractional quantum anomalous Hall
effect in the star lattice. The $\nu=-1$ plateau also comes from
the gap opened at Dirac point. But this gap will disappear when
$\frac{t_2}{t_1}$ increase. The reason that $\nu=-1$ can appear in
our model is that the bands of the star lattice (\figref{fig2})
are narrower than those of honeycomb lattice and kagome lattice.
This results in that a direct gap can be opened above the lowest
band. So, when $\frac{t_2}{t_1}$ increases, the bands become
wider, the direct gap become an indirect gap, the Hall
conductivity becomes not quantized.

\begin{figure}
\begin{center}
\includegraphics[width=8.5cm]{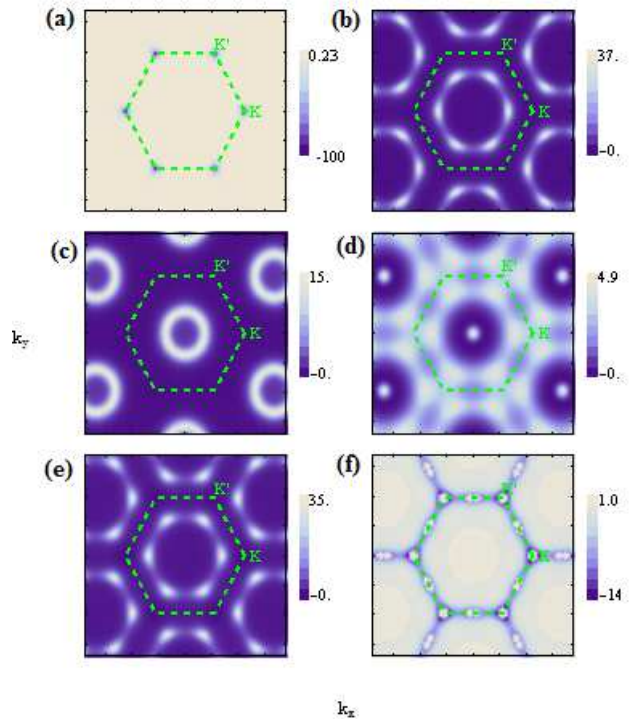}
\end{center}
\caption{The total Berry curvature distribution in momentum space
$\Omega_{z}\left(k\right)=\sum_{E_n (\leq
E_f)}\Omega_{z}\left(E_n,k\right)$ for the star lattice with
$t_2=t_1$, $t_{RSO}=0.2t_1$, $\lambda=0.2 t_1$, when Fermi level
lies at (a) $E_f=-2.77$. (b) $E_f=-2.32$. (c) $E_f=-0.21$. (d)
$E_f=0.36$. (e) $E_f=1.27$. (f) $E_f=1.0$}\label{fig4}
\end{figure}

To further confirm the above interpretation, we plot the total
Berry curvature of the bands below the Fermi level
$\Omega_{z}\left(k\right)=\sum_{E_n (\leq
E_f)}\Omega_{z}\left(E_n,k\right)$. When the Fermi level lies in
the gaps, peaks at the minimal energy difference points are
clearly seen (\figref{fig4}).

\begin{figure}
\begin{center}
\includegraphics[width=8.5cm]{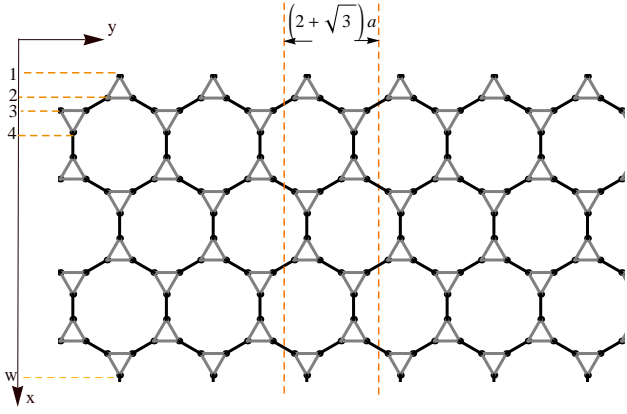}
\end{center}
\caption{Illustration of a zigzag star lattice ribbon which is
infinite in $y$ direction and has width $W$ in $x$ direction. The
unit cell is indicated by the dashed lines.}\label{fig5}
\end{figure}

\begin{figure}
\begin{center}
\includegraphics[width=8.5cm]{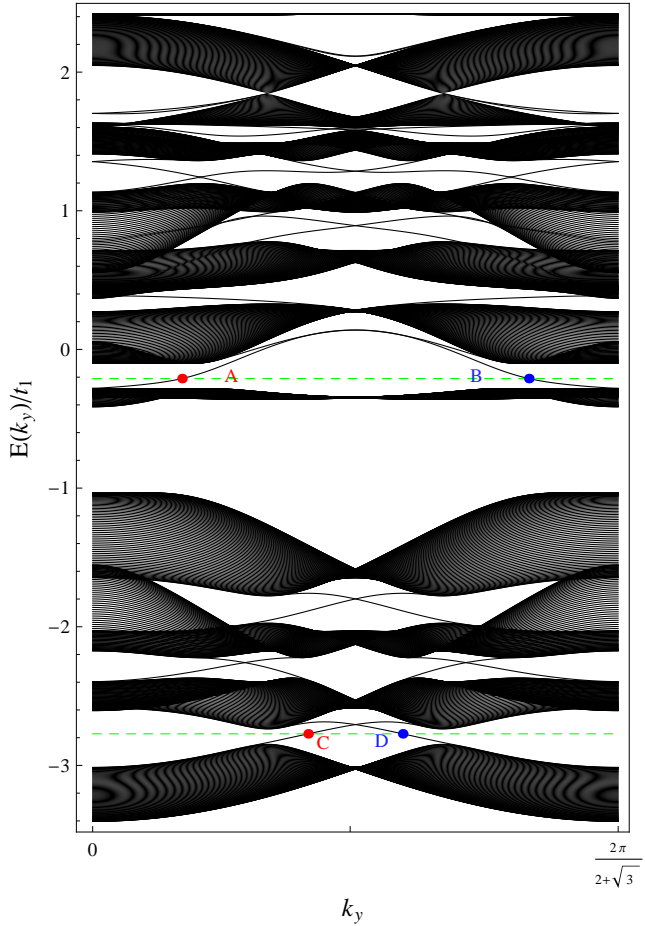}
\end{center}
\caption{(Color online) The band structure of a W=324 width zigzag
star lattice ribbon with $t_2=t_1$, $t_{RSO}=0.2t_1$, $\lambda=0.2
t_1$. Edge bands appear in the topological nontrival
gaps.}\label{fig6}
\end{figure}

{\it Chiral Edge States}.--The nonzero Chern number also manifests
as the presence of chiral edge states which localize at edge and
propagate in one direction. Now, we study the edge state property
on the zigzag star lattice ribbon (\figref{fig5}). Periodic
boundary condition is taken along $y$ direction (parallel to the
edge). Open boundary condition is taken along $x$ direction. The
band structure of a 324-sites width ribbon is shown in
\figref{fig6}.


We can see that edge bands appear in topological gaps, and no edge
bands present in the gap above the fourth band. The absence of
edges bands in this gap confirms that it is a topological trivial
gap, as mentioned.

\begin{figure}
\begin{center}
\includegraphics[width=8.5cm]{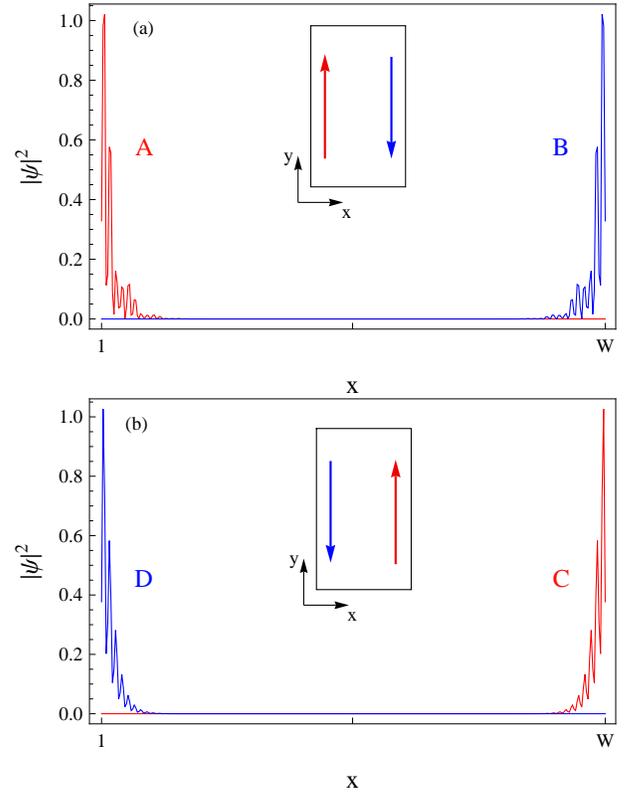}
\end{center}
\caption{(Color online) The edge states probability for Fermi
level indicated in \figref{fig6} (green dashed lines). (a)
$E_f=-0.21 t_1$ and $k_y=\frac{17}{100}\frac{2\pi}{(2+\sqrt{3})a}$
for the point marked as A,
$k_y=\frac{83}{100}\frac{2\pi}{(2+\sqrt{3})a}$ for point B. (b)
$E_f=-2.77 t_1$ and $k_y=\frac{41}{100}\frac{2\pi}{(2+\sqrt{3})a}$
for the point marked as C,
$k_y=\frac{59}{100}\frac{2\pi}{(2+\sqrt{3})a}$ for point
D.}\label{fig7}
\end{figure}

To further demonstrate the topological property of these gaps, we
calculate the edge state functions corresponded to the points
marked on the edge bands (\figref{fig6}). The probability density
quickly decrease as going into the bulk. The group velocity
$\frac{\partial E(k)}{\partial k}$ suggests these edge states
propagate in different directions at different edge
(\figref{fig7}). So chiral edge states appear when Fermi level
lies in the topological nontrival bulk gaps. The number of these
edge states indicate the absolute value of Chern number. The
chiral of the edge states indicate the sign of Chern number.

{\it Conclusion}.--In summary, we have studied the quantum
anomalous Hall effect on star lattice in the presence of both
Rashba spin-orbit effect and exchange field. The Chern number
calculated from the Berry curvature in momentum space and the edge
states calculated on the zigzag ribbon confirm the emergence of
QAH effect when Fermi level lies in the topological gaps. Due to
the complex lattice structure of star lattice, our results include
what are obtained in honeycomb lattice which has Chern number
$\nu=2$ and kagome lattice which has Chern number
$\nu=1~\text{or}~2$. What's more, a nearly flat band with nonzero
Chern number ($\nu=1$) also appear in our model, which provide the
great potential to realize the fractional quantum Hall effect
without tuning the ration between nearest and next-nearest hopping
amplitude \cite{Wen2011,SunKai2011,Mudry2011}.

Finally, the physical realization of the star lattice would be not
very difficult considering its two cousins, honeycomb lattice and
kagome lattice, have been realized in optical lattices
\cite{Duan2003,Ruostekoski2009}. We also notice that the
spin-orbit coupling has also been realized in optical lattice
\cite{Spielman2011}. Both of these offer the possibility to
realize QAH effect in the star optical lattice. In solid state
system, a ploymeric iron(III) acetate with underlying star lattice
has also been reported \cite{Long2007}.

{\it Acknowledgements}.--We thank Liang Chen and Yuanpei Lan for
helpful discussions. This work was supported by the NSFC under
Grant No.10675108.


\end{document}